\documentclass{PoS}

\def\beqn{\begin{eqnarray}} \def\eeqn{\end{eqnarray}}

\title{Increasing the precision for $Z$ production at colliders: mixed QCD-QED effects}

\ShortTitle{Increasing the precision for $Z$ boson production}

\author{\speaker{German F. R. Sborlini}$^{\ a,b}$\\
        $^a$Instituto de F\'{\i}sica Corpuscular, Universitat de Val\`{e}ncia -- 
Consejo Superior de Investigaciones Cient\'{\i}ficas, Parc Cient\'{\i}fic, E-46980 Paterna, Valencia, Spain.\\
 		$^b$Dipartimento di Fisica, Universit\`a di Milano and INFN Sezione di Milano,
I-20133 Milan, Italy.\\
        E-mail: \email{german.sborlini@unimi.it}}

\abstract{In this talk, we describe the recent progress on the inclusion of mixed QCD-QED corrections for collider observables. In particular, we developed a formalism to extend $q_T$-resummation to deal with simultaneous emission of gluons and photons. We applied it to $Z$ production at colliders, and briefly discuss extensions to more complicated final states.}

\FullConference{
European Physical Society Conference on High Energy Physics - EPS-HEP2019 -\\
			10-17 July, 2019\\
			Ghent, Belgium}

\begin{document}

\section{Introduction and motivation}
\label{sec:introduction}
During the recent years, the need for more precise theoretical predictions has become crucial for the progress of high-energy physics. Up to now, the Standard Model (SM) showed an impressive agreement with the experimental data, within the estimated error bands available. However, the increasing precision of the experiments forces to reduce theoretical uncertainties, since many new physics phenomena might be hided within any tiny discrepancy.

In this talk, we center the discussion on the inclusion of higher-order corrections to the production of vector bosons in hadronic collisions. In fact, the Drell-Yan (DY) \cite{Drell:1970wh} process is often considered as the standard candle to extract highly accurate data and perform a very precise comparison with the available theoretical models. Since hadronic colliders are dominated by QCD interactions, the natural refinement of the theoretical predictions was based on the computation of higher-orders within perturbative QCD. In this way, the next-to-leading order (NLO) corrections were first obtained in the seventies \cite{Drell:1970wh,Altarelli:1978id}, whilst the next-to-next-to-leading order (NNLO) was calculated by several groups since the nineties \cite{Hamberg:1990np,Anastasiou:2003yy}. 

However, an accurate phenomenological description of this process requires to properly deal with soft radiation originated from the colliding particles. This can be taken into account by resumming logarithmically-enhanced contributions, for instance using the $q_T$-resummation formalism \cite{Catani:2013tia}. In fact, this formalism has been succesfully applied to compute up to next-to-next-to-leading logarithmic (NNLL) QCD corrections for DY \cite{Bozzi:2008bb,Bozzi:2010xn,Catani:2015vma}. Alternative methods have been developed to reach up to N$^3$LL+NNLO accuracy \cite{Catani:2014uta,Bizon:2018foh}, which are the current state-of-the-art in precision QCD calculations for DY.

On top of resummation and QCD corrections, the presence of electroweak (EW) particles in intermediate and final states might have a non-negligible impact on the phenomenological description of several processes. For instance, we explored higher-order QED effects to diphoton production \cite{Sborlini:2017gpl,Sborlini:2018fhr} and mixed QCD-QED contributions to DGLAP equations \cite{deFlorian:2015ujt,deFlorian:2016gvk,Sborlini:2016dfn}. In both cases, we found tiny effects however the introduction of QED terms led to a noticeable reduction of the uncertainties related to the EW scheme choice. In the case of DY, higher-order EW corrections should be included to provide a completely consistent calculation \cite{Wackeroth:1996hz}. Very recently, mixed QCD-QED and higher-order QED corrections to this process were studied by including the fixed-order terms \cite{deFlorian:2018wcj,Delto:2019ewv}. 

The purpose of this talk consists in combining fixed-order QCD-QED corrections within a proper generalization of the resummation formalism. We based our strategy on the $q_T$-resummation framework and the Abelianization algorithm, which has been succesfully applied to study QCD-QED corrections to the Altarelli-Parisi splitting functions \cite{deFlorian:2015ujt,deFlorian:2016gvk}. In the following, we will briefly describe the key ingredients of the formalism, recalling some useful formulae of the $q_T$-resummation/subtraction method in Sec. \ref{sec:qTresummation}. Then, in Sec. \ref{sec:QCDQEDresumed}, we will enter into the details of the Abelianization and the extension of the formalism. Also, we will present some explicit results. Finally, the conclusions and future research lines are depicted in Sec. \ref{sec:conclusions}.

\section{Removing IR divergences: $q_T$-resummation formalism}
\label{sec:qTresummation}
The $q_T$-resummation/subtraction formalism turns out to be very important to capture the phenomenological impact of soft radiation from the initial state \cite{Catani:2013tia}. Let's consider an arbitrary colorless final state $F$ produced in hadronic collisions, and let's consider the transverse-momentum relative to the collision axis, $\vec{q}_T$, of this final state. The singular contribution to the differential cross-section can be expressed as 
\beqn
\nonumber \frac{d \sigma_{F+X}}{d^2 \vec{q}_T \, dM} &=& \frac{M^2}{s} \sum_{c=\{q,\bar{q},g\}} \, \left[d\hat{\sigma}^{(0)}_{c \bar{c} \to F} \right] \, \int \frac{d^2 \vec{b}}{4\pi^2} \, e^{\imath \vec{b}\cdot\vec{q}_T} \, S_c(M,b) \, 
\\ &\times& \sum_{a_1,a_2} \, \int_{x_1}^1 \frac{dz_1}{z_1}\, \int_{x_2}^1 \frac{dz_2}{z_2} \, \left[H^V C_1 C_2 \right]_{a_1 a_2 \to c \bar{c}} \, f_{a_1}^{h_1}(x_1/z_1,b_0^2/b^2) f_{a_2}^{h_2}(x_2/z_2,b_0^2/b^2) \, ,
\label{eq:Master1}
\eeqn
where $d\hat{\sigma}^{(0)}_{c \bar{c} \to F}$ is the leading-order partonic cross-section, $f_{a_i}^{h_j}(x,Q)$ are the PDFs associated to the density distribution of a parton $a_i$ inside an hadron $h_i$, $[H^V C_1 C_2]$ is the \textit{hard-collinear} factor and $S_c$ is the Sudakov factor corresponding to the soft/collinear gluon emission from a parton $c$. In this formula, all these components are process-independent except for the LO cross-section and the \textit{hard-virtual} coefficient $H^V$. This last coefficient encodes the information relative to the virtual amplitudes, after a proper removal of the IR singularities through the application of the subtraction operators defined in Ref. \cite{Catani:1998bh}.

Another important detail about the master expression given in Eq. (\ref{eq:Master1}) is related to the $b$-space formulation. In particular, it is worth appreciating that the PDFs are evaluated at the reference scale $b_0^2/b^2$, which involves including an additional routine within the code to perform the DGLAP evolution. The advantage of this approach is that the resummed cross-section can be computed separately, since it is given by
\beqn
\frac{d \hat{\sigma}_{a + b \to F}^{\rm res}}{d q_T^2}(q_T,M) &=& \frac{M}{\hat{s}} \, \int_0^\infty db \, \frac{b}{2} \, J_0(b \, q_T) {\cal W}_{ab}(b,M,\hat{s}) \, ,
\label{eq:Master2}
\\({\cal W}_{ab})_N &=& \hat{\sigma}_{a+b \to F}^{(0)} \, {\cal H}^F_N(\alpha_S(\mu_R),\mu_R,\mu_F,Q^2) \, \exp\{{\cal G}_N(\alpha_S,L,\mu_R,Q) \} \, ,
\label{eq:Master3}
\eeqn
where all the logarithmic terms are contained inside ${\cal W}_{ab}$, which can be decomposed in the Mellin space. Explicitly, $\exp\{{\cal G}_N\}$ is a universal form factor and ${\cal H}^F$ is the hard-virtual contribution to the $N$-th Mellin momenta of ${\cal W}_{ab}$. It is worth appreciating that the \textit{hard} terms in Eqs. (\ref{eq:Master1}) and (\ref{eq:Master2}) can be connected through some specific tranformations, as explained in Ref. \cite{Catani:2013tia}. However, they contain the explicit process-dependence of the virtual matrix elements, which involve that they must be computed for each process separately.

\section{Mixed QCD-QED resummation}
\label{sec:QCDQEDresumed}
As we motivated in the Introduction, EW corrections play a crucial role within the precision physics program. Thus, we have to properly estimate the contributions due to multiple and simultaneous emissions of soft gluons and photons. This requires to extend the $q_T$-resummation/subtraction formalism to combine the QCD and QED effects in a consistent way. In Ref. \cite{Cieri:2018sfk}, we use Eq. (\ref{eq:Master2}) as starting point, and we recover the pure QED version. By applying the Abelianization algorithm \cite{deFlorian:2015ujt,deFlorian:2016gvk}, we manage to consider the multiple soft-photon emission and obtain a similar description of QED resummation as the one provided by the traditional YFS formalism \cite{Yennie:1961ad}. In fact, it is worth appreciating that the Abelian nature of QED makes it easier to resum logarithmically-enhanced contributions due to the lack of non-trivial correlations among particles.

The next step consisted in keeping track of the simultaneous emission of soft gluons and photons, and extend the validity of the master formula given in Eq. (\ref{eq:Master2}). We noticed that, in the soft/collinear region, the kinematical behaviour of photons and gluons was similar, thus the assumptions involved in the derivation of the $q_T$-resummation approach were still valid. There were two important features to take in account:
\begin{itemize}
\item the Mellin-space formulation, with the decomposition of the form factor ${\cal W}_{ab}$ as presented in Eq. (\ref{eq:Master3});
\item and the \textit{mixed QCD-QED} renormalization group equations. 
\end{itemize}
The last point was crucial to properly define the ${\cal G}$ functions inside the transformed form factor ${\cal W}$. In fact, these equations are:
\beqn
\frac{d \ln{\alpha_S(\mu^2)}}{d \ln{\mu^2}} &=& - \sum_{n=0}^\infty \beta_n \, \left(\frac{\alpha_S}{\pi}\right)^{n+1} - \sum_{m=1,n=0}^\infty \beta_{n,m} \, \left(\frac{\alpha_S}{\pi}\right)^{n+1} \left(\frac{\alpha}{\pi}\right)^{m} , \
\label{eq:RUNNINGqcd}
\\ \frac{d \ln{\alpha(\mu^2)}}{d \ln{\mu^2}} &=& - \sum_{n=0}^\infty \beta'_n \, \left(\frac{\alpha}{\pi}\right)^{n+1} - \sum_{m=1,n=0}^\infty \beta'_{n,m} \, \left(\frac{\alpha}{\pi}\right)^{n+1} \left(\frac{\alpha_S}{\pi}\right)^{m} , \
\label{eq:RUNNINGqed}
\eeqn
where $\beta_{n,m}$ and ${\beta'}_{n,m}$ are the beta-functions coefficients in a double-perturbative expansion for QCD and QED, respectively. Explicitly, we computed the first non-trivial mixing terms \cite{Cieri:2018sfk}, which are given by
\beqn
\beta_{0,1} = - \frac{1}{8} \, \sum_{q=1}^{n_f} e_q^2 \, , \ \ \ \ && \ \ \ \ {\beta'}_{1,0} = - \frac{C_A \, C_F}{8} \, \sum_{q=1}^{n_f} e_q^2 \, .
\eeqn
We need to consider this coupled and mixed evolution because it originates the non-trivial QCD-QED mixing terms in the logarithmic expansion of the Sudakov form factor. Explicitly, by using this strategy we can generalize Eq. (\ref{eq:Master3}) by defining
\beqn
\nonumber {\cal G}'_{N}(\alpha_S,\alpha,L) &=& {\cal G}_{N}(\alpha_S,L) \, + \,  L \, g'^{(1)}(\alpha L) + \sum_{n=2}^\infty \, \left(\frac{\alpha}{\pi}\right)^{n-2} \, g'^{(n)}(\alpha L) 
\\ &+& \sum_{n,m=1}^{\infty} \, \left(\frac{\alpha_S}{\pi}\right)^{n-2}\left(\frac{\alpha}{\pi}\right)^{m-2} g'^{(n,m)}(\alpha_S L, \alpha L)  \, ,
\label{eq:Gprime}
\\ {\cal H}'^F_N(\alpha_S,\alpha) &=& {\cal H}^F_N(\alpha_S) + \sum_{n=1}^{\infty} \, \left(\frac{\alpha}{\pi}\right)^n {\cal H}'^{F\, (n)}_N + \sum_{n,m=1}^{\infty} \, \left(\frac{\alpha_S}{\pi}\right)^n \left(\frac{\alpha}{\pi}\right)^m\,  {\cal H}'^{F\, (n,m)}_N \, ,
\label{eq:Hprime}
\eeqn
where ${\cal H}'^F_N$ and $\exp\{{\cal G'}_N(\alpha_S,\alpha,L,\mu_R,Q^2)\}$ are the hard-collinear and resummed factors, respectively. In these formulae, we appreciate three different contributions: pure QCD (unprimed terms), pure QED (primed single-indexed terms) and non-trivially mixed QCD-QED contributions (primer double-indexed). As in the standard QCD formalism, the hard-collinear part is process-dependent, but the $g$-functions are universal. In particular, we obtained
\beqn
\nonumber g'^{(1,1)} &=& \frac{{A'}^{(1)}_q\, \beta_{0,1}}{\beta_0^2 \, \beta'_{1,0}} \left[\ln(1-\lambda') \, \left(\frac{\lambda(1-\lambda')}{(1-\lambda)(\lambda-\lambda')}+\ln\left(\frac{\lambda'(1-\lambda)}{\lambda'-\lambda}\right)\right) - \frac{\lambda'}{\lambda-\lambda'}\ln(1-\lambda) \right.
\\ &-& \left. {\rm Li}_2\left(\frac{\lambda}{\lambda-\lambda'}\right)+{\rm Li}_2\left(\frac{\lambda(1-\lambda')}{\lambda-\lambda'}\right) \right] \, + ({\rm primed}\leftrightarrow {\rm unprimed}) \, ,
\eeqn
with $\lambda=\beta_0 \alpha_s \, L$ and $\lambda'=\beta'_0 \alpha \, L$, being $L$ the \emph{large-logarithm}. This corresponds to the first non-trivial QCD-QED mixed logarithmic correction\footnote{More details about the notation used and the derivation of the formalism is available in Ref. \cite{Cieri:2018sfk}.}.

\begin{figure}[htb]
\begin{center}
\begin{tabular}{cc}
\includegraphics[width=0.42\textwidth]{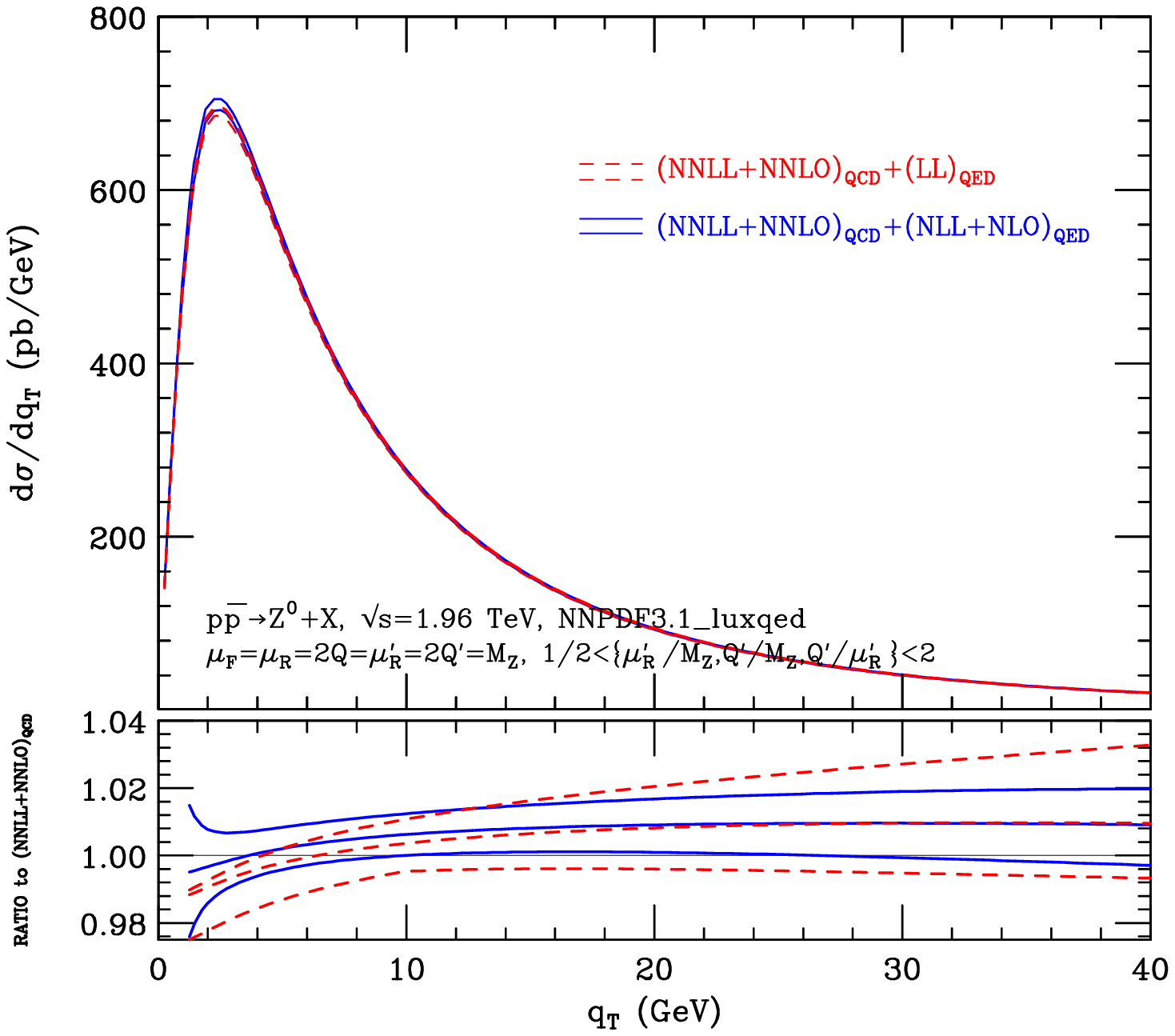} & \ \ \includegraphics[width=0.42\textwidth]{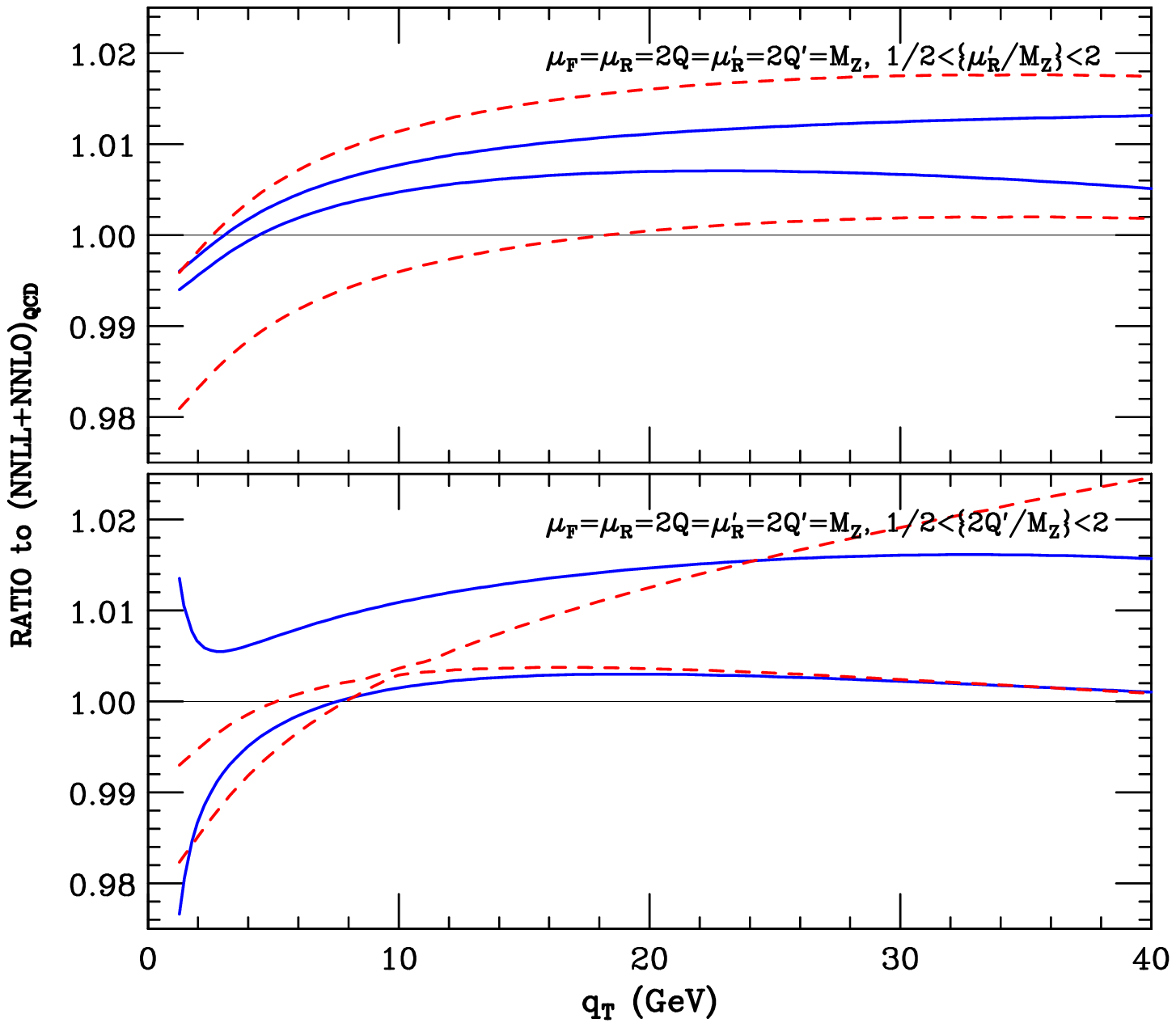}\\
\end{tabular}
\end{center}
\caption{\label{fig:figura1}
Higher-order QCD-QED corrections to $Z$-boson production at Tevatron, with NNLL+NNLO QCD predictions as default (black lines). In the left panel, we include LL (red dashed) and NLL+NLO (blue solid) QED corrections. We plot the ratio of these corrections compared to the QCD default prediction, as well as the corresponding error bands. The uncertainty bands when varying the resummation (upper plot) and renormalization (lower plot) QED scales are presented in the right panel.}
\end{figure}

\subsection{Application to $Z$-boson production}
\label{ssec:Zbosonapp}
In order to test the combined resummation formalism, we applied it to compute the mixed QCD-QED corrections to $Z$-boson production at colliders. As an example, we considered the case of the $q_T$ spectrum of the $Z$-boson being produced on-shell at Tevatron ($E_{CM}=1.96$ TeV). For the computational setup, we used the \texttt{NNPDF3.1LUXqed} PDF set \cite{Manohar:2017eqh}, and the usual\footnote{Note that within this formalism, we can freely choose the renormalization/resummation scales for QCD/QED independently.} central-scale choice $\mu_F=\mu_R=2 Q=m_Z$. The results are shown in Fig. \ref{fig:figura1}, where the reference prediction corresponds to NNLL+NNLO QCD in the narrow-width approximation. On top of that, we included LL QED (red dashed lines) and \textit{NLL'+NLO QED} (blue solid lines) corrections. Here, \textit{$NLL'$} stands for taking into account the non-trivial mixing terms in both the $g$-functions and the evolution of the running couplings.

As we can appreciate from the plots, the effects are non-negligible (i.e. percent-level close to the resummation peak) but small compared with pure QCD corrections. However, the importance of adding these higher-order contributions is the reduction the scale uncertainties, thus making the predictions more stable with respect to the choice of the EW parameters and scheme \cite{Sborlini:2018fhr,Cieri:2018sfk,INPREP}.

%%%%%%%%%%%%%%%%%%%%%%%%%%%%%%%%%%%%%%%%%%%%%%%%%%%%
\section{Outlook and conclusions}
\label{sec:conclusions}
In this article, we briefly explained the extension of the $q_T$-resummation/subtraction formalism to deal with mixed QCD-QED corrections. We based the strategy on the application of the Abelianization algorithm, to first obtain the pure QED formalism and then proceed to the consistent extension for simultaneously tackling QCD-QED radiation.

As a phenomenological example, we considered the $Z$-boson production in the narrow-width approximation. In Ref. \cite{Cieri:2018sfk} we analyzed the $q_T$ spectrum at LHC and Tevatron, and we found sub-percent level deviations from the pure NNLL+NNLO QCD predictions. Of course, in the sight of future improvements in the experimental precision, these corrections are non-negligible. Most importantly, the inclusion of higher-order mixed QCD-QED terms improves the perturbative stability of the predictions when varying the EW parameters, thus contributing to a more reliable result.

Due to the theoretical importance of a consistent treatment of higher-order mixed corrections, we are currently investigating an extension to deal with charged final states (i.e. $W^{\pm}$ production \cite{INPREP}). Moreover, any advance in this direction could benefit the computation of precise predictions for any QFT involving more than two interactions, even beyond the SM.

\section*{Acknowledgments}
\label{sec:Acknowledgements}
This work is supported by the Spanish Government (Agencia Estatal de Investigacion) and ERDF funds from European Commission (Grants No. FPA2017-84445-P and SEV-2014-0398), by Generalitat Valenciana (Grant No. PROMETEO/2017/053), by Consejo Superior de Investigaciones Cient\'ificas (Grant No. PIE-201750E021) and Fondazione Cariplo under the Grant No. 2015-0761. The author also acknowledges the support by the COST Action CA16201 PARTICLEFACE.

\end{document}